\documentclass[camera,letterpaper,nomarginnotes,nonarrowgutter,hyperref]{jpaper}

\usepackage{cite}
\usepackage{amsmath,amssymb,amsfonts}
\usepackage{graphicx}
\usepackage{textcomp}
\usepackage{xcolor}
\usepackage{fancyhdr}
\usepackage{hyperref}
\usepackage{titlesec}
\usepackage{flushend}
\usepackage{multirow}
\usepackage{tabularx}
\usepackage{booktabs}
\usepackage{setspace}
\usepackage{caption}
\usepackage{array}
\usepackage{algorithm}
\usepackage[noend]{algpseudocode}
\algrenewcommand\algorithmicindent{1em}
\makeatletter\newcommand{\LState}[1]{\State\parbox[t]{\dimexpr\linewidth-\ALG@thistlm}{#1\strut}}\makeatother
\AtBeginDocument{\emergencystretch=3em}

\newcommand{\MAEReq}{\mathrm{MAE}_{\mathrm{Req}}}
\newcommand{\CoreReduction}{56.5}
\newcommand{\RequestReduction}{7.4}
\usepackage{tcolorbox}
\usepackage{tikz}

\newcommand{\secref}[1]{\S\ref{#1}}
\newcommand{\figref}[1]{Fig.~\ref{#1}}
\newcommand{\figsref}[1]{Figs.~\ref{#1}}
\newcommand{\tabref}[1]{Table~\ref{#1}}
\renewcommand{\algref}[1]{Alg.~\ref{#1}}
\newcommand{\head}[1]{\noindent\textbf{#1.}~}

\DeclareRobustCommand{\circled}[1]{\tikz[baseline=(c.base)]\node[circle,fill={rgb,255:red,24;green,61;blue,75},text=white,inner sep=0pt,minimum size=2.7mm,font=\sffamily\bfseries\scriptsize] (c) {#1};}

\newcounter{obsv}\newcounter{takeaway}
\definecolor{boxaccent}{RGB}{23,97,122}
\newtcolorbox{obsvbox}{colback=boxaccent!6!white, colframe=boxaccent, boxrule=.65pt,
  sharp corners, boxsep=0pt, left=5pt, right=5pt, top=3.2pt, bottom=3.2pt, before skip=6pt, after skip=6pt}
\newtcolorbox{takeawaybox}{colback=boxaccent!14!white, colframe=boxaccent, boxrule=.65pt,
  sharp corners, boxsep=0pt, left=5pt, right=5pt, top=3.5pt, bottom=3.5pt, before skip=6.5pt, after skip=6.5pt}
\newcommand{\observation}[1]{\refstepcounter{obsv}\par\begin{obsvbox}{\color{boxaccent}\textbf{Observation~\theobsv.}}~#1\end{obsvbox}\noindent\ignorespaces}
\newcommand{\takeaway}[1]{\refstepcounter{takeaway}\par\begin{takeawaybox}{\color{boxaccent}\textbf{Takeaway~\thetakeaway.}}~#1\end{takeawaybox}}

\begin{document}
\title{Synthesizing Fast and Accurate Immediate-Response\\DRAM Latency Models with Agents}

\author{Haocong Luo
\quad Nisa Bostanc{\i}
\quad Ataberk Olgun
\quad Harsh Songara
\quad Jikun Wang \\
Konstantinos Sgouras
\quad Konstantinos Kanellopoulos
\quad Andreas Kosmas Kakolyris
\quad Onur Mutlu \\[0.45em]
{SAFARI Research Group}
\\
ETH Zurich}

\hypersetup{pdftitle={Synthesizing Fast and Accurate Immediate-Response DRAM Latency Models with Agents},
  pdfauthor={Haocong Luo, Nisa Bostanci, Ataberk Olgun, Harsh Songara, Jikun Wang, Konstantinos Sgouras, Konstantinos Kanellopoulos, Andreas Kosmas Kakolyris, Onur Mutlu}}

\maketitle

\begin{abstract}
Accurately modeling DRAM latency is critical for computer architecture simulation. Cycle-level DRAM simulators are accurate but slow and hard to integrate into fast and parallel simulators, which therefore use immediate-response DRAM models that predict each request's latency once at admission. Unfortunately, existing immediate-response models rely on simplified analytical formulas that abstract away row-buffer locality, request reordering, and write draining, and manually designing accurate models across DRAM standards is laborious.

Our goal is to develop an agentic methodology in which large language model (LLM) agents autonomously synthesize immediate-response DRAM latency models that are fast, accurate, and white-box (i.e., readable and explainable). To this end, we build a development harness around Ramulator~2 and ChampSim in which an agent iteratively proposes a model, evaluates it against a cycle-level oracle on both final core-cycle count and per-request latency, diagnoses errors using request traces and synthetic traffic, and refines the model.

We conduct ten-iteration synthesis runs with four LLMs. On 24 test workloads, the most accurate model (Opus~5.5) achieves 0.15\% mean core-cycle error and 0.063 normalized per-request error, which are \CoreReduction$\times$ and \RequestReduction$\times$ lower than the most accurate baseline on each metric (Mess at 8.57\% and queued FixedLat at 0.465), at 6.4$\times$ standalone DRAM speedup over the oracle. The fastest model (GPT-6 Astra) achieves 1.54\% core-cycle error and 0.201 per-request error (5.6$\times$ and 2.3$\times$ lower) at 31.6$\times$ speedup. Our analysis shows that the models are accurate on read traffic below saturation, while writes remain a major error source: the models cannot predict when buffered writes issue, and three of the four account for writes upon arrival. Without modification, the models retain their accuracy in gem5's out-of-order core (0.30--1.12\% vs.\ 2.99--4.43\% elapsed-time error for baselines) and, for three of the four models, across nine DRAM and controller configurations, while accelerating four-core zsim multiprogram simulation by 3.6--7.3$\times$. On LPDDR6 and HBM4, new synthesis runs reach 0.04--0.58\% core-cycle error and unmodified DDR5 models maintain 0.65--1.07\%, versus at least 14.03\% for every baseline. We will open-source our harness and synthesized models.
\end{abstract}

\section{Introduction}
\label{sec:introduction}
Accurately modeling DRAM latency is critical for the fidelity of computer architecture simulation. Cycle-level DRAM simulators (e.g., Ramulator~2~\cite{kim2016ramulator,ramulatorgithub,luo2024ramulator2,luo2026ramulator21,ramulator2github} and DRAMSim3~\cite{li2020dramsim3}) are accurate because they model every DRAM command, timing constraint, and scheduling decision cycle by cycle. Unfortunately, this makes them slow and hard to use in fast and parallel simulators. For example, zsim~\cite{sanchez2013zsim} runs 13--15$\times$ slower with Ramulator or DRAMSim3 than with an immediate-response DRAM model~\cite{esmailidokht2024mess}, and integrating a cycle-level DRAM simulator into Sniper~\cite{carlson2011sniper} requires extensive modifications to its core model~\cite{eyerman2021irram}.

Fast and parallel simulators therefore use \emph{immediate-response} DRAM models, which predict a request's latency once at admission and never revise it (\secref{sec:bg-models}). Existing immediate-response models use simplified analytical formulas, such as a queueing model~\cite{sanchez2013zsim,carlson2011sniper} or a measured bandwidth--latency curve (Mess~\cite{esmailidokht2024mess}). Unfortunately, such formulas abstract away the behavior that causes DRAM latency to vary substantially across requests, e.g., row-buffer hits and conflicts, bank contention, request reordering, and write draining.

Designing, evaluating, and optimizing an immediate-response model that is both accurate and fast is laborious. Human engineers must manually inspect DRAM request traces, test diverse traffic patterns to identify latency-determining mechanisms, implement them in low-overhead admission-time code, carefully examine their interactions, and repeat this effort for each new DRAM standard. Because this kind of work is repetitive and straightforward to verify against a cycle-level oracle, it is well suited to LLM agents.

\textbf{Our goal} is to develop an agentic methodology that synthesizes fast, accurate, and white-box (generic, configurable, and auditable) immediate-response DRAM latency models. To this end, we build a development harness based on CHIA~\cite{cui2026chia}, in which an LLM agent iteratively proposes a model, evaluates it against a cycle-level oracle (ChampSim~\cite{gober2022champsim} with Ramulator~2), diagnoses errors with request traces and synthetic traffic, and refines it (\secref{sec:methodology}). \textbf{The key idea} is to evaluate each proposed model on both final core-cycle count and per-request latency: fitting only average latency can achieve low core-cycle error through error cancellation, whereas optimizing per-request error guides the agent to model the underlying DRAM mechanisms. An independent LLM session reviews every candidate, and the agents never see the test workloads.

We conduct ten-iteration synthesis runs with four LLMs on single-core DPC-4 traces~\cite{dpc4}. All four synthesized models outperform all four baselines on both metrics (\secref{sec:single-core-results}). On 24 test workloads, the most accurate model (Opus~5.5) achieves 0.15\% mean core-cycle error and 0.063 normalized per-request error, which are \CoreReduction$\times$ and \RequestReduction$\times$ lower than the most accurate baseline on each metric (Mess at 8.57\% and queued FixedLat at 0.465, respectively), at 6.4$\times$ DRAM speedup over the oracle. The fastest model (GPT-6 Astra) achieves 1.54\% core-cycle error and 0.201 per-request error (5.6$\times$ and 2.3$\times$ lower) at 31.6$\times$ speedup.

Our model analysis yields three key findings. First, the agents synthesize readable, auditable code whose accuracy is determined by a few interpretable mechanisms (\secref{sec:ablation}). Second, controlled synthetic sweeps show that the models are accurate on read traffic below saturation, but fail on writes, on FR-FCFS reordering (which no immediate-response model can predict), and near saturation (\secref{sec:failure-modes}). Third, writes cause major residual error in two ways: the models cannot predict when buffered writes issue (the 18.9\% of test reads overlapping a write account for 33--47\% of per-request error), and three of the four models account for writes upon arrival, over-predicting subsequent reads (\secref{sec:failure-modes}).

Without modification, the synthesized models retain their accuracy advantage in gem5's out-of-order core and, except for one model with hard-coded parameters, across nine DRAM and controller configurations (\secref{sec:extensions}). On LPDDR6 and HBM4, new synthesis runs reach 0.04--0.58\% core-cycle error and GPT-6 Astra's and Opus~5.5's unmodified DDR5 models maintain 0.65--1.07\%, versus at least 14.03\% for every baseline (\secref{sec:standard-transfer}). In zsim, the models accelerate four-core multiprogram mixes by 3.6--7.3$\times$ over cycle-level Ramulator~2 (\secref{sec:parallel-simulators}).

This paper makes the following key contributions:
\begin{itemize}
    \item To our knowledge, this is the first work to use LLM agents to synthesize immediate-response DRAM latency models. We propose an agentic methodology for developing fast, accurate immediate-response models that can extend to other system components.
    \item We introduce a closed-loop development harness in which LLM agents synthesize, diagnose, and optimize DRAM latency models against a cycle-level oracle (\secref{sec:methodology}).
    \item We synthesize models that are 1)~more accurate than existing baselines (up to \CoreReduction$\times$ and \RequestReduction$\times$ lower core-cycle and per-request error), 2)~fast (6.4--31.6$\times$ standalone DRAM speedup), 3)~configurable across DRAM organizations and controllers, 4)~human-auditable, and 5)~applicable across DDR5, LPDDR6, and HBM4 (\secref{sec:single-core-results}, \secref{sec:extensions}, \secref{sec:parallel-simulators}).
    \item We identify the interpretable mechanisms and failure modes that explain the accuracy and speed of the synthesized models (\secref{sec:model-analysis}).
    \item We will open-source our development harness, synthesized models, and evaluation infrastructure.
\end{itemize}

\section{Background and Motivation}
\label{sec:background}

We review DRAM and controller behavior (\secref{sec:bg-dram}, \secref{sec:bg-scheduling}), cycle-level versus immediate-response models (\secref{sec:bg-models}), and our motivation (\secref{sec:bg-motivation}).

\begin{figure*}[t]
    \centering
    \includegraphics[width=\textwidth]{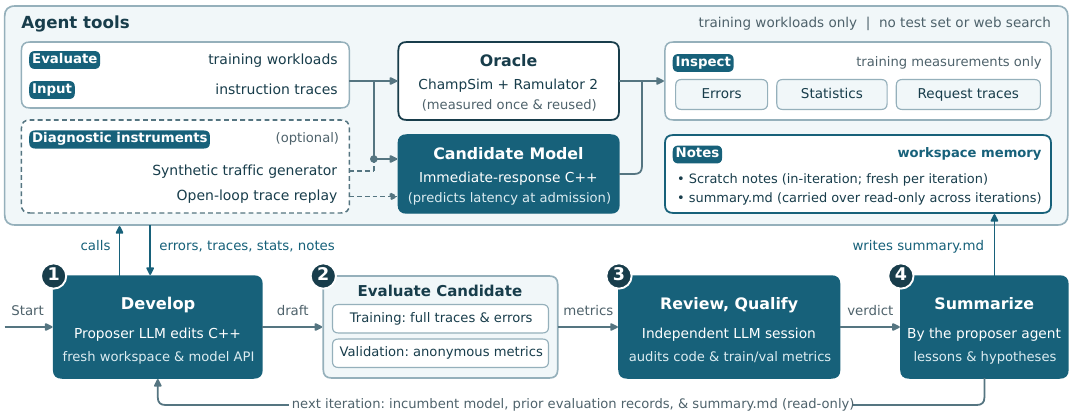}
    \caption{Workflow of our development harness. Top: six agent tools (teal) on training workloads only (no test set or web search), including \textbf{Notes} for in-iteration scratch notes and cross-iteration summaries (\texttt{summary.md}); diagnostic instruments (dashed) are optional. Bottom: one iteration of the synthesis loop (\circled{1}--\circled{4}).}
    \label{fig:workflow}
\end{figure*}

\subsection{DRAM Organization and Operation}
\label{sec:bg-dram}
A DRAM channel connects the memory controller to one or more ranks, each composed of bank groups and banks (8 bank groups $\times$ 4 banks in our baseline, \tabref{tab:simulation-configuration}). Each bank is a row array with a \emph{row buffer} holding the most recently opened row. To access a cache line, the controller issues 1)~an activate (\texttt{ACT}) command to open a row, 2)~a read (\texttt{RD}) or write (\texttt{WR}) command to transfer data, and 3)~a precharge (\texttt{PRE}) command to close the row. Latency therefore depends on bank state: a \emph{row hit} requires only \texttt{RD}/\texttt{WR}, a closed bank requires \texttt{ACT} first, and a \emph{row conflict} requires \texttt{PRE} and \texttt{ACT} first.

DRAM standards define inter-command timing constraints~\cite{jedec2020ddr5}, including, e.g., $t_{\mathrm{RCD}}$ (\texttt{ACT}-to-\texttt{RD}/\texttt{WR}), $t_{\mathrm{RP}}$ (\texttt{PRE}-to-\texttt{ACT}), $t_{\mathrm{RC}}$ (\texttt{ACT}-to-\texttt{ACT}), $t_{\mathrm{CCD\_S}}$/$t_{\mathrm{CCD\_L}}$ (column commands across/within bank groups), $t_{\mathrm{RRD}}$ (\texttt{ACT} rate limit), $t_{\mathrm{WR}}$ (write recovery, \texttt{WR}-to-\texttt{PRE}), $t_{\mathrm{RTP}}$ (\texttt{RD}-to-\texttt{PRE}), and read/write bus turnarounds (e.g., $t_{\mathrm{WTR}}$). Banks serve requests in parallel but share command and data buses.

\begin{table}[t]
    \centering
    \caption{ChampSim and Ramulator~2 configuration.}
    \label{tab:simulation-configuration}
    \begingroup\footnotesize\setlength{\tabcolsep}{4pt}\renewcommand{\arraystretch}{1.08}
\begin{tabularx}{\columnwidth}{@{}lX@{}}\toprule
\textbf{Component} & \textbf{Configuration} \\\midrule
Core & Out-of-order, 4 GHz \\
Fetch/decode/dispatch & \mbox{6/6/6/4/5} instructions/cycle \\
\quad /execute/retire &  \\
ROB; load/store queue & 352; 128/72 entries \\
Branch prediction & Bimodal; basic BTB \\
L1I & 32 KiB, 8-way, 4 cycles \\
L1D & 48 KiB, 12-way, 5 cycles \\
L2 & 512 KiB, 8-way, 10 cycles \\
LLC & 2 MiB, 16-way, 20 cycles \\
Cache line; prefetching & 64 B; disabled \\
Pages; first-touch penalty & 4 KiB; 200 cycles \\
Physical placement & Timing-independent, preassigned \\
Warmup / measured & 2M / 20M instructions \\
\midrule
DRAM & DDR5-4800AN, 8 GiB, $\times$8 devices \\
Channel/rank/banks & 1 / 1 / $8\times4$ (groups $\times$ banks) \\
Oracle; row policy & GenericDDR / FRFCFS-RowHit; open \\
Read/write queues & 64/64 entries \\
Write-drain low/high & 0.5/0.8 \\
Refresh & Disabled \\
\bottomrule\end{tabularx}\endgroup

\end{table}

\subsection{Request Scheduling and Write Draining}
\label{sec:bg-scheduling}
The memory controller buffers incoming requests in read and write queues and selects one command per cycle. The widely used FR-FCFS policy~\cite{rixner2000memory} prioritizes row-buffer hits and then older requests, improving throughput but reordering accesses so a request's latency depends on requests arriving \emph{after} it.

Because writes (e.g., cache writebacks) are off the core's critical path once accepted by the controller, the controller buffers writes and prioritizes reads. To amortize bus turnarounds, it drains writes in batches using watermarks: our oracle enters \emph{write mode} when the write queue exceeds its high watermark (80\% of 64 entries) or when the read queue is empty, and returns to read mode only when the write queue drops below its low watermark (50\%) and a read waits. Reads arriving during write mode wait; hence a read's latency can depend on earlier buffered writes and whether the read queue was empty when a drain started.

\subsection{Cycle-Level and Immediate-Response DRAM Models}
\label{sec:bg-models}
A cycle-level model (e.g., Ramulator~2~\cite{kim2016ramulator,ramulatorgithub,luo2024ramulator2,luo2026ramulator21,ramulator2github}) simulates the controller and DRAM cycle by cycle, holding each request until its final command completes. Unlike a cycle-level model, an immediate-response model predicts a request's latency at admission and returns it immediately using only past state, never revising it. This contract makes immediate-response models fast and compatible with simulators whose components do not advance in lockstep, such as zsim's bound/weave phases~\cite{sanchez2013zsim} and Sniper's loosely synchronized cores~\cite{carlson2011sniper}, but makes high accuracy challenging under FR-FCFS reordering and write draining.

\subsection{Motivation}
\label{sec:bg-motivation}
Two observations motivate our work. First, cycle-level DRAM simulation overhead matters most in fast and parallel simulators. With a detailed sequential out-of-order core, DRAM simulation takes 10--30\% of runtime in gem5 O3CPU with DRAMSim3, rising to 70--80\% with a simpler core~\cite{li2019thesis}. In parallel simulators, cycle-level DRAM dominates: zsim with Ramulator or DRAMSim3 runs 13--15$\times$ slower than with Mess~\cite{esmailidokht2024mess}, and DRAMSim2 caps zsim at about 3~MIPS~\cite{sanchez2013zsim}. Second, existing immediate-response models leave a large accuracy gap: in our evaluation (\secref{sec:single-core-results}), FixedLat, M/D/1, and WMG1 incur 8.95--10.54\% mean core-cycle error (up to 55.2\% max), and Mess incurs 8.57\% (up to 52.9\% max).

This accuracy gap exists because existing immediate-response models rely on coarse-grained aggregate traffic statistics and are fundamentally \emph{address-agnostic}. For example, Mess~\cite{esmailidokht2024mess} models DRAM latency by indexing random-probe bandwidth--latency curves using aggregate bandwidth and read/write ratio. Prior work~\cite{crowthers2026panem} finds that these curves become multi-valued when DRAM is oversaturated and that Mess does not capture the latency changes on transient, short bursts of memory traffic. Another work~\cite{baek2026address} shows that Mess fails to capture DRAM latency variations as memory-request addresses change, especially when DRAM is \emph{not} fully saturated. Ignoring request addresses misses large latency variations from row-buffer locality (up to 36.4\%), read/write ordering (up to 21.3\%), and inter-channel skewness (up to 4.7$\times$)~\cite{baek2026address}.

We conclude that closing this accuracy gap requires immediate-response DRAM models that track address- and state-dependent behavior at low overhead, and an automated methodology to synthesize them without repetitive manual effort.

\section{Development Harness and Evaluation Methodology}
\label{sec:methodology}

We build a closed-loop development harness based on CHIA~\cite{cui2026chia} enabling LLM agents to iteratively propose, evaluate, and optimize an immediate-response DRAM latency model. Below, we present the workflow and model interface, followed by the oracle and agent tools (\secref{sec:agent-tools}), metrics (\secref{sec:metrics}), workloads (\secref{sec:workloads}), qualification (\secref{sec:qualification}), synthesis runs (\secref{sec:runs}), and baselines (\secref{sec:baselines}).

\figref{fig:workflow} shows the workflow across four steps. First, the agent develops a candidate model~\circled{1} using agent tools (\secref{sec:agent-tools}) on training workloads as often as it chooses. Second, the harness evaluates the candidate~\circled{2} on visible training workloads and anonymous validation workloads (\secref{sec:workloads}). Third, an independent LLM session reviews the candidate~\circled{3} and decides whether it replaces the incumbent (\secref{sec:qualification}). Finally, the agent writes a summary~\circled{4} of what it tried and learned for the next iteration.

\head{Model Interface} A synthesized model is a C++ class in Ramulator~2 with one admission-time prediction function that immediately returns a request's completion cycle. At initialization, the model receives the hardware description (DRAM organization, all timing parameters, read/write queue capacities, and write watermarks) and its parameter file. The controller retains each request until its predicted completion cycle and applies backpressure when queues are full (matching the oracle). The task specification requires that every model 1)~predicts each latency once at admission, 2)~never revises an issued prediction, 3)~uses no future or internal oracle state, 4)~does not schedule commands cycle by cycle, and 5)~does not specialize to workload names or addresses.

\subsection{Oracle and Agent Tools}
\label{sec:agent-tools}
Our harness wraps Ramulator~2~\cite{kim2016ramulator,ramulatorgithub,luo2024ramulator2,luo2026ramulator21,ramulator2github} (coupled with ChampSim~\cite{gober2022champsim}) with six agent tools invoked during model development~\circled{1} (and summarization~\circled{4}):
\begin{enumerate}
    \item \textbf{Input.} Exposes the ChampSim instruction trace of any training workload.
    \item \textbf{Evaluate.} Runs the candidate on all training workloads against the oracle to measure modeling accuracy.
    \item \textbf{Inspect.} Returns DRAM request traces (coordinates, read/write type, arrival/departure cycles) and simulation statistics for error diagnosis.
    \item \textbf{Synthetic Traffic Generator.} Synthesizes custom DRAM access streams at the agent's discretion to isolate hypotheses and failure modes.
    \item \textbf{Open-Loop Replay.} Feeds the oracle's recorded request arrival timestamps through the candidate without closed-loop feedback.
    \item \textbf{Notes.} Allows every agent to read and write scratch notes and the required end-of-iteration summary~\circled{4}; each iteration starts in a fresh workspace, and only prior summaries, evaluation records, and candidate models carry forward across iterations.
\end{enumerate}
\tabref{tab:simulation-configuration} summarizes our ChampSim and Ramulator~2 configuration.

\subsection{Key Accuracy Metrics}
\label{sec:metrics}
To ensure models capture true DRAM and scheduling behavior rather than canceling errors across requests, agents optimize two metrics jointly: 1)~final core-cycle error $E_C$, and 2)~the relative latency error of each \emph{individual} DRAM request ($\MAEReq$, matched between the model and oracle). For model and oracle core cycles $C_m, C_o$ and matched request latencies $\ell_{m,i}, \ell_{o,i}$ ($i \in P$),
\begin{equation}
\begin{aligned}
 E_C &= \frac{|C_m-C_o|}{C_o} \times 100\%, \\
 \MAEReq &= \frac{\sum_{i\in P}|\ell_{m,i}-\ell_{o,i}|}{|P|L},
\end{aligned}
\end{equation}
where $L$ is the mean latency over all recorded oracle requests.

\subsection{Workloads}
\label{sec:workloads}
We construct 14 training, 14 validation, and 24 test workloads from DPC-4 traces~\cite{dpc4}, spanning Google server traces, SPEC CPU2017~\cite{bucek2018spec}, GAP~\cite{beamer2015gap}, GMS~\cite{besta2021graphminesuite}, Ligra~\cite{shun2013ligra}, and AI/ML workloads (\tabref{tab:workloads}). During candidate evaluation~\circled{2}, training workloads are fully visible, validation exposes only per-workload error metrics under anonymous aliases, and test workloads remain completely hidden from the harness and agents.

\begin{table}[t]
    \centering
    \caption{Single-core DPC-4 workload split (52 workloads total).}
    \label{tab:workloads}
    \begingroup\footnotesize\setlength{\tabcolsep}{3pt}\renewcommand{\arraystretch}{1.12}
\begin{tabularx}{\columnwidth}{@{}l*{3}{>{\raggedright\arraybackslash}X}@{}}\toprule
 & \textbf{Train (14)} & \textbf{Validation (14)} & \textbf{Test (24)} \\\midrule
Google & arizona, bravo, charlie, whiskey & delta, merced, tango, yankee & sierra (7), tahoe (8) \\
SPEC17 & cactuBSSN, lbm, mcf, omnetpp & bwaves, fotonik3d, xz & cam4, gcc, pop2, xalancbmk \\
Graph & GAP bfs, sssp; GMS kstar, maxclique & GAP pr; Ligra CC; GMS subgraph, triangle & GAP bc; Ligra CF, radii \\
AI/ML & Llama2-7B, Whisper & RWKV, SD-v1.4, ViT-L & Bark, BioGPT \\
\bottomrule\end{tabularx}\endgroup

\end{table}

\subsection{Qualification of Proposed Models}
\label{sec:qualification}
In qualification~\circled{3}, an independent LLM session reviews the candidate code and train/validation metrics ($E_C$ and $\MAEReq$) to verify 1)~compliance with the immediate-response contract, and 2)~whether metric changes merit promotion. Using an LLM reviewer rather than a rigid scalar score permits meaningful gains on one metric (or on anonymous validation) despite minor regressions elsewhere.

\subsection{Synthesis Runs and LLMs}
\label{sec:runs}
We conduct four ten-iteration synthesis runs on single-core workloads with four LLMs: 1)~Gemini 3.8 Flash, 2)~DeepSeek V4.1 Flash, 3)~GPT-6 Astra, and 4)~Opus~5.5 (\secref{sec:single-core-results}). All four agents run without web search, access the same six tools (including \textbf{Notes}), and start each iteration in a fresh workspace carrying forward only prior summaries, evaluation records, and candidate sources. \secref{sec:extensions} evaluates generalization without modification to gem5's out-of-order core (\secref{sec:frontend-transfer}) and alternative DRAM organizations/queue depths (\secref{sec:dram-transfer}), evaluates LPDDR6 and HBM4 using both unmodified DDR5 models and four additional ten-iteration synthesis runs with GPT-6 Astra and Opus~5.5 (\secref{sec:standard-transfer}), and presents preliminary multicore synthesis runs (\secref{sec:multicore}).

\subsection{Baselines}
\label{sec:baselines}
We compare against four untuned immediate-response baselines: 1)~queued FixedLat (constant access latency plus bandwidth queueing), 2)~zsim's M/D/1~\cite{sanchez2013zsim}, 3)~Sniper's windowed M/G/1 (WMG1)~\cite{carlson2011sniper}, and 4)~Mess~\cite{esmailidokht2024mess,messgithub} (measured bandwidth--latency curves). We measure Mess's curves on our oracle and set queued FixedLat, M/D/1, and WMG1 to match the oracle's zero-load read latency (110 DRAM cycles, 45.8~ns).

\section{Foundational Single-Core Results}
\label{sec:single-core-results}

We evaluate the four models synthesized from ten-iteration single-core runs using Gemini 3.8 Flash (high), DeepSeek V4.1 Flash (max), GPT-6 Astra (xhigh), and Opus~5.5 (xhigh) on accuracy (\secref{sec:sc-accuracy}), standalone speedup (\secref{sec:sc-speedup}), LLM cost (\secref{sec:llm-usage}), and modeled mechanisms (\secref{sec:insights}).

\subsection{Simulation Accuracy}
\label{sec:sc-accuracy}

\begin{figure}[t]
    \centering
    \includegraphics[width=\columnwidth]{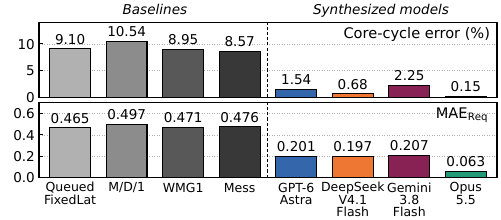}
    \caption{Mean core-cycle error (top) and $\MAEReq$ (bottom) across the 24 single-core DDR5-4800 test workloads.}
    \label{fig:foundation-accuracy}
\end{figure}

\begin{figure*}[t]
    \centering
    \includegraphics[width=\textwidth]{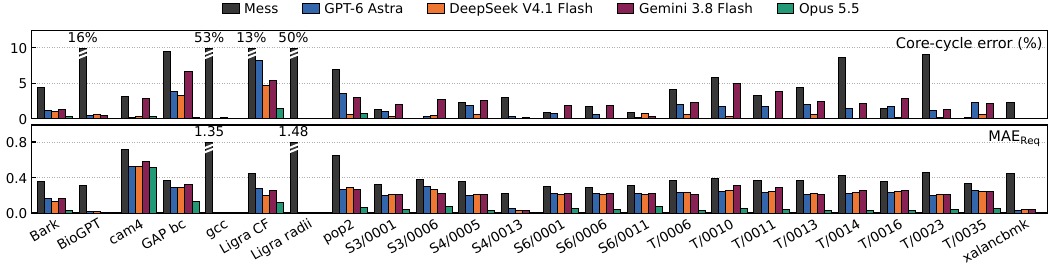}
    \caption{Per-workload test-set errors (core-cycle, top; $\MAEReq$, bottom) for Mess and the four synthesized models. S$g$/$r$ and T/$r$ denote Google sierra (group $g$, region $r$) and tahoe (region $r$).}
    \label{fig:per-workload}
\end{figure*}

\figref{fig:foundation-accuracy} compares test-set mean core-cycle error (top) and $\MAEReq$ (bottom) across the 24 test workloads for all four baselines (left) and synthesized models (right), and \figref{fig:per-workload} breaks down per-workload errors for Mess and the four synthesized models.

\observation{\label{obs:sc-accuracy}All four synthesized models outperform all four baselines on both mean core-cycle error and mean $\MAEReq$.}
We highlight three results from \figsref{fig:foundation-accuracy} and~\ref{fig:per-workload}. First, Opus~5.5 achieves the lowest errors: 0.15\% mean core-cycle error (1.53\% max on Ligra CF) and 0.063 mean $\MAEReq$ (0.516 max on cam4), which are \CoreReduction$\times$ lower than the best baseline on core-cycle error (Mess, at 8.57\% mean and 52.92\% max) and \RequestReduction$\times$ lower than the best baseline on $\MAEReq$ (queued FixedLat, at 0.465 mean and 1.478 max). Second, DeepSeek V4.1 Flash, GPT-6 Astra, and Gemini 3.8 Flash achieve 0.68\% (4.71\% max), 1.54\% (8.23\% max), and 2.25\% (6.65\% max) mean core-cycle error, and 0.197, 0.201, and 0.207 mean $\MAEReq$, respectively. Third, Opus~5.5 beats Mess on core-cycle error across all 24 workloads (vs.\ 22 for DeepSeek V4.1 Flash, 21 for GPT-6 Astra, and 16 for Gemini 3.8 Flash), and all four models beat Mess on $\MAEReq$ across all 24 workloads.

\begin{figure}[t]
    \centering
    \includegraphics[width=\columnwidth]{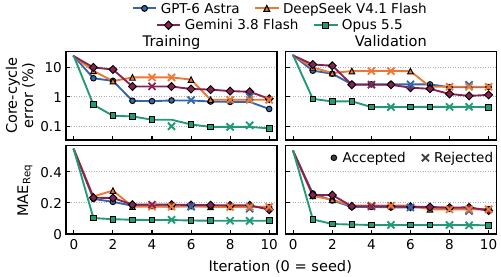}
    \caption{Training (left) and validation (right) error evolution (core-cycle, top, log scale; $\MAEReq$, bottom) over ten synthesis iterations.}
    \label{fig:foundation-convergence}
\end{figure}

\figref{fig:foundation-convergence} plots training (left) and validation (right) error trajectories across the ten iterations (x-axis; iteration~0 is the seed model), with circles and crosses marking accepted and rejected candidates. Additional iterations generally improve accuracy across all LLMs, though convergence rates differ: Opus~5.5 reaches 0.85\% validation core-cycle error after iteration~1, whereas Gemini 3.8 Flash still remains at 1.17\% after iteration~10.

\subsection{Simulation Speedup}
\label{sec:sc-speedup}

\begin{figure}[t]
    \centering
    \includegraphics[width=\columnwidth]{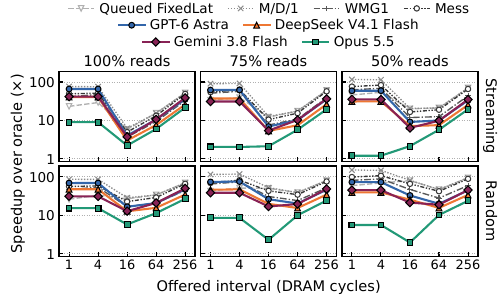}
    \caption{Standalone DRAM simulation speedup over the cycle-level oracle by pattern, read ratio, and offered arrival interval (median of five runs).}
    \label{fig:foundation-speed}
\end{figure}

To isolate DRAM model speed, we replace ChampSim with an open-loop synthetic traffic generator generating streaming and random requests at three read ratios and five fixed arrival intervals (backpressure may delay admission without reordering). We replay identical streams (100K warmup, 5M measured requests) through each model and the oracle on an Intel Granite Rapids core. \figref{fig:foundation-speed} plots standalone speedup over the oracle (y-axis, log scale) across arrival intervals (x-axis), patterns (rows), and read fractions (columns).

\observation{\label{obs:sc-speedup}Synthesized models achieve 6.4--31.6$\times$ geometric-mean standalone speedup over the oracle, with GPT-6 Astra approaching analytical baseline speeds.}
Across all 30 points, geometric-mean speedups are 31.63$\times$ (GPT-6 Astra), 23.60$\times$ (Gemini 3.8 Flash), 21.73$\times$ (DeepSeek V4.1 Flash), and 6.44$\times$ (Opus~5.5), versus 32.47--54.22$\times$ for the baselines. Opus~5.5 exhibits strong traffic sensitivity: at a one-cycle interval with 50\% reads, its speedup is 1.20$\times$ for streaming versus 5.56$\times$ for random traffic, rising to 19.38$\times$ at a 256-cycle streaming interval.

\subsection{LLM Usage and Cost}
\label{sec:llm-usage}

\tabref{tab:llm-usage} reports token usage and estimated API cost\footnote{GPT-6 Astra and Opus~5.5 ran via subscriptions; API-equivalent costs are estimated from recorded tokens and provider pricing.} for each ten-iteration synthesis run.

\begin{table}[t]
    \centering
    \caption{Token usage (output includes reasoning) and API-equivalent cost per ten-iteration synthesis run. $^\dagger$Gemini 3.8 Flash did not fully report cache hit rate on all input tokens, so its reported cost spans 0--100\% hit rate.}
    \label{tab:llm-usage}
    \begingroup\footnotesize\setlength{\tabcolsep}{4pt}\renewcommand{\arraystretch}{1.1}
\begin{tabular*}{\columnwidth}{@{\extracolsep{\fill}}lrrr@{}}\toprule
\textbf{LLM} & \textbf{Input (M)} & \textbf{Output (M)} & \textbf{API cost (\$)} \\\midrule
GPT-6 Astra & 92.8 & 0.46 & 639.03 \\
DeepSeek V4.1 Flash & 195.6 & 3.05 & 7.98 \\
Gemini 3.8 Flash & 583.2 & 1.70 & 57.85--73.34$^\dagger$ \\
Opus~5.5 & 252.8 & 3.35 & 209.87 \\
\bottomrule\end{tabular*}\endgroup

\end{table}

\observation{\label{obs:llm-cost}Synthesis cost varies by 80$\times$ across LLMs, and an inexpensive LLM (\$7.98) synthesizes a highly accurate model.}
Input volume spans 92.8M--583.2M tokens, output spans 0.46M--3.35M tokens, and estimated cost spans \$7.98 (DeepSeek V4.1 Flash) to \$639.03 (GPT-6 Astra). DeepSeek V4.1 Flash achieves 2.3$\times$ lower core-cycle error than GPT-6 Astra (0.68\% vs.\ 1.54\%) and comparable $\MAEReq$ (0.197 vs.\ 0.201) at roughly 80$\times$ lower API cost. All agents make extensive use of diagnostic tools: across ten iterations, GPT-6 Astra, DeepSeek V4.1 Flash, Gemini 3.8 Flash, and Opus~5.5 invoke 186, 32, 48, and 208 synthetic sweeps and 85, 59, 307, and 34 open-loop replays, respectively, using synthetic traffic in every iteration.

\takeaway{Agentic synthesis produces white-box models that trade off accuracy and speed, achieving 3.8--\CoreReduction$\times$ lower core-cycle error and 2.2--\RequestReduction$\times$ lower $\MAEReq$ than the best baseline on each metric, at 6.4--31.6$\times$ standalone DRAM speedup and as low as \$8 in API cost.}

\subsection{Insights from Representative Models}
\label{sec:insights}

\begin{table*}[t]
    \centering
    \caption{Summary of modeled mechanisms, code size, DDR5-4800 test accuracy, and standalone speedup across all four single-core synthesized models.}
    \label{tab:model-mechanisms}
    \begingroup\footnotesize\setlength{\tabcolsep}{5pt}\renewcommand{\arraystretch}{1.3}
\begin{tabularx}{\textwidth}{@{}>{\raggedright\arraybackslash}p{0.145\textwidth}*{4}{>{\raggedright\arraybackslash}X}@{}}
\toprule
\textbf{Attribute} & \textbf{GPT-6 Astra} & \textbf{DeepSeek V4.1 Flash} & \textbf{Gemini 3.8 Flash} & \textbf{Opus~5.5} \\
\midrule
\multicolumn{5}{@{}l}{\textit{Modeled mechanisms}}\\[1pt]
\textbf{Row locality} & Last admitted row; per-type rows; 8 recent rows under backlog & Per-bank open row & Per-bank open row and \texttt{PRE}/\texttt{ACT} readiness & 8 open-row windows per bank (row hits first) \\
\rowcolor{black!6}\textbf{Contention} & Free-time clocks per bank, bank group, and channel; activation clocks & First-fit data-bus calendar (128-entry scan) & Burst calendar plus a fitted congestion polynomial & First-fit \texttt{ACT} and column calendars with bounded column slip \\
\textbf{Writes} & Accounted for upon arrival; write-recovery and write-to-read floors; amortized turnaround & Booked at admission; fitted conflict penalty & FIFO write queue; opportunistic per-read drain and high-watermark drain (247 cycles at 64-entry capacity) & Replayed write mode: watermarks, FR-FCFS drain, forwarding \\
\rowcolor{black!6}\textbf{DRAM traffic prediction} & None & None & None & Next-read and writeback forecasts \\
\midrule
\multicolumn{5}{@{}l}{\textit{Implementation}}\\[1pt]
\textbf{Configurability} & Reads hardware description & Hard-coded DDR5-4800, 32 banks, and address decode & Reads hardware description (64-entry write-queue cap) & Reads hardware description \\
\rowcolor{black!6}\textbf{Fitted constants} & None (\texttt{\{\}}) & 3 (write penalties) & 6 (congestion, drain rate) & 1 + 10 mode switches \\
\textbf{C++ lines of code} & 191 & 246 & 390 & 783 \\
\midrule
\multicolumn{5}{@{}l}{\textit{Results}}\\[1pt]
\textbf{DDR5 test error} & 1.54\% $E_C$, 0.201 $\MAEReq$ & 0.68\% $E_C$, 0.197 $\MAEReq$ & 2.25\% $E_C$, 0.207 $\MAEReq$ & 0.15\% $E_C$, 0.063 $\MAEReq$ \\
\rowcolor{black!6}\textbf{Standalone speedup} & 31.6$\times$ & 21.7$\times$ & 23.6$\times$ & 6.4$\times$ \\
\bottomrule
\end{tabularx}\endgroup

\end{table*}

\tabref{tab:model-mechanisms} summarizes the modeled mechanisms, code complexity, accuracy, and standalone speedup of all four synthesized models. All four maintain lightweight per-resource state and compute each request's latency once at admission without cycle-by-cycle simulation. Unlike baselines that estimate queueing from aggregate load, all four track individual DRAM resource availability, ranging from 191 to 783 physical lines of C++ (from $O(1)$ free-time clocks with zero fitted parameters to first-fit command calendars with replayed write drains). Between the speed and accuracy endpoints, DeepSeek V4.1 Flash pairs per-bank row state with a 128-entry first-fit data-bus calendar and a fitted $+$90-cycle post-write conflict penalty (reaching 0.68\% $E_C$, but hard-coding DDR5-4800 geometry and address bits), whereas Gemini 3.8 Flash combines per-bank readiness, a burst calendar, a FIFO write queue with opportunistic and high-watermark drains, and a fitted congestion polynomial (2.25\% $E_C$). Below, we detail the fastest (GPT-6 Astra) and most accurate (Opus~5.5) models.

\head{GPT-6 Astra: Resource Availability Clocks} GPT-6 Astra maintains free-time clocks per bank ($b.\mathit{ready}$), bank group ($G_g$), and channel ($C$), plus activation clocks ($t_{\mathrm{RRD\_S}}$, $t_{\mathrm{RRD\_L}}$), predicting latency via \algref{alg:astra}. At arrival $t$, it computes row-state delay $\rho$, bank service time $s$, activation wait $A$, and total wait $W$, updates the resource clocks (adding burst length $n_{\mathrm{BL}}$ on $C$), and returns $\lceil t_{\mathrm{RL}}+1+W \rceil$. All timings come from the hardware description; constant work per admission yields its 31.6$\times$ speedup.

\begin{figure*}[t]
\begin{minipage}[t]{0.488\textwidth}
\begin{algorithm}[H]
\caption{GPT-6 Astra's latency prediction (simplified). Symbols: $\rho$ row delay, $s$ bank service time, $A$ activation wait, $W$ total wait, $C$/$G_g$ channel/bank-group clocks, $n_{\mathrm{BL}}$ burst length.}
\label{alg:astra}
\footnotesize
\begin{algorithmic}[1]
\Function{Access}{bank $b$, bank group $g$, row $r$, cycle $t$}
\State $\rho \gets 0 \mid t_{\mathrm{RCD}} \mid t_{\mathrm{RCD}}+t_{\mathrm{RP}}-1$ \Comment{hit / closed / conflict}
\LState{$\mathit{hit} \gets r$ is $b.\mathit{row}$, $b$'s last row of this type or, under backlog, one of $b$'s 8 recent rows}
\State $s \gets t_{\mathrm{CCD\_L}}$ \textbf{if} $\mathit{hit}$ \textbf{else} $\max(t_{\mathrm{RC}}-\mathit{run},\, t_{\mathrm{RTP}}+t_{\mathrm{RP}}+t_{\mathrm{RCD}})$
\LState{\textbf{if} $b$'s last access was a write \textbf{then} raise $s$ to a ramped write-recovery (conflict) or write-to-read (same-row read) floor}
\LState{$A \gets 0$ \textbf{if} $\mathit{hit}$ \textbf{else} the wait on the activation clocks ($t_{\mathrm{RRD\_S}}$, $t_{\mathrm{RRD\_L}}$)}
\State $W \gets \max(t+\rho+A,\; b.\mathit{ready}+s,\; C,\; G_g) - t$
\State $b.\mathit{ready} \gets t+W$; \ $b.\mathit{row} \gets r$
\State $C \gets \max(t+\rho,\, C) + n_{\mathrm{BL}}$, plus a turnaround for a write
\State $G_g \gets \max(t,\, G_g) + t_{\mathrm{CCD\_L}}$
\State \Return latency $\lceil t_{\mathrm{RL}}+1+W \rceil$
\EndFunction
\end{algorithmic}
\end{algorithm}
\end{minipage}\hfill
\begin{minipage}[t]{0.488\textwidth}
\begin{algorithm}[H]
\caption{Opus~5.5's latency prediction (simplified). Symbols: $\mathit{WQ}$ write queue, $n_{\mathrm{CL}}$ CAS latency, $n_{\mathrm{BL}}$ burst length.}
\label{alg:opus}
\footnotesize
\begin{algorithmic}[1]
\Function{Write}{$w$, cycle $t$}
\State \Call{Advance}{$t$}; append $w$ to the write queue $\mathit{WQ}$; \Return $t+1$
\EndFunction
\Function{Read}{$r$, cycle $t$}
\State \Call{Advance}{$t$}
\State \textbf{if} $r$ matches a write in $\mathit{WQ}$ \textbf{then return} $t+1$ \Comment{forwarding}
\State \textbf{if} in write mode \textbf{then} drain $\mathit{WQ}$ below the low watermark
\State predict the next read's arrival from recent inter-arrival gaps
\LState{book \texttt{PRE}, \texttt{ACT}, and \texttt{RD} on the bank and bus calendars at the first fit, inside an open-row window if \texttt{RD} fits before its \texttt{PRE}}
\LState{\textbf{if} $r$'s \texttt{ACT} empties the read queue \textbf{then} commit buffered writes before \texttt{RD}, each adding a write-to-read turnaround}
\State \Return $\mathit{RD}+n_{\mathrm{CL}}+n_{\mathrm{BL}}$
\EndFunction
\Function{Advance}{$t$}
\LState{replay write mode up to $t$; each step commits the earliest-issuable buffered write (FR-FCFS)}
\EndFunction
\end{algorithmic}
\end{algorithm}
\end{minipage}
\end{figure*}

\head{Opus~5.5: Command Calendars and Internal Write Scheduler} Opus~5.5 synthesizes a model structurally similar to IRRam~\cite{eyerman2021irram} (\secref{sec:related-work}): it schedules \texttt{PRE}, \texttt{ACT}, and \texttt{RD} using per-bank timers and channel-wide \texttt{ACT}/column calendars, serves row hits within open-row windows, and defers writes into drains (\algref{alg:opus}, returning $\mathit{RD}+n_{\mathrm{CL}}+n_{\mathrm{BL}}$, where $n_{\mathrm{CL}}$ is CAS latency). Unlike Gemini 3.8 Flash's fixed-length drain, Opus~5.5 replays its write queue via an event-driven FR-FCFS scheduler~\cite{rixner2000memory} governed by the oracle's watermarks (lines~6 and~12), augmented with online next-read and dirty-writeback forecasts. This write scheduler explains much of Opus~5.5's superior accuracy and part of its lower speed; its calendar-based read path also makes it 3.0$\times$ slower than GPT-6 Astra even on read-only traffic.

\section{Analysis of the Synthesized Models}
\label{sec:model-analysis}

A white-box model is useful only if architects understand what it computes (\secref{sec:insights}), which mechanisms matter (\secref{sec:ablation}), and where it fails (\secref{sec:failure-modes}). We focus on GPT-6 Astra (fastest) and contrast it with Opus~5.5 (most accurate).

\subsection{Ablation Study}
\label{sec:ablation}
To identify which of GPT-6 Astra's mechanisms determine accuracy, we remove one mechanism at a time (editing 1--5 lines of C++ source) and re-evaluate on the 24 test workloads (\tabref{tab:astra-ablation-full}). We make four key observations.

\begin{table}[t]
    \centering
    \caption{Mechanism-removal ablation of GPT-6 Astra on the 24 single-core DDR5-4800 test workloads.}
    \label{tab:astra-ablation-full}
    \begingroup\footnotesize\setlength{\tabcolsep}{4pt}\renewcommand{\arraystretch}{1.08}
\begin{tabular*}{\columnwidth}{@{\extracolsep{\fill}}>{\raggedright\arraybackslash}p{0.62\columnwidth}rr@{}}\toprule
\textbf{Change to GPT-6 Astra's model} & \begin{tabular}[b]{@{}r@{}}\textbf{Core-cycle}\\\textbf{error (\%)}\end{tabular} & \textbf{$\MAEReq$} \\\midrule
None (control) & 1.54 & 0.201 \\\midrule
Remove all per-bank state & 12.23 & 0.428 \\
Remove the open-row proxy (every access pays $t_{\mathrm{RCD}}$) & 4.19 & 0.473 \\
Remove the write-recovery floor & 4.03 & 0.219 \\
Remove the channel clock & 2.28 & 0.188 \\
Fix the channel clock's row-opening delay & 2.23 & 0.187 \\
Remove the amortized write turnaround & 1.58 & 0.200 \\
Remove the write-to-read floor & 1.54 & 0.201 \\
Remove the activation throttles & 1.54 & 0.201 \\
Remove the bank-group clock & 1.54 & 0.201 \\
Remove the recent-row history under backlog & 1.51 & 0.209 \\
Remove the last row per request type & 0.65 & 0.216 \\
\bottomrule\end{tabular*}\endgroup

\end{table}

\observation{\label{obs:ablation-bank}Per-bank state is essential to prediction accuracy.} Removing all per-bank state raises mean core-cycle error from 1.54\% to 12.23\% and $\MAEReq$ from 0.201 to 0.428. Without bank backlog tracking, the model under-predicts bank-bound workloads, worsening signed core-cycle error from $-3.8\%$ to $-30.4\%$ on GAP bc; without row-hit tracking, it over-predicts streaming workloads, raising signed error from $-0.01\%$ to $+24.6\%$ on gcc.

\observation{\label{obs:ablation-metrics}Core-cycle and per-request accuracy respond differently to individual mechanisms.} Removing the open-row proxy degrades both metrics, more than doubling $\MAEReq$ (0.201 to 0.473, worse on all 24 workloads) and raising core-cycle error to 4.19\% (2.7$\times$), even with partial opposite-sign error cancellation between row hits and conflicts. Removing the write-recovery floor affects core-cycle accuracy almost exclusively (raising core-cycle error to 4.03\%, while $\MAEReq$ rises only to 0.219), most strongly on write- and conflict-heavy workloads where signed error worsens from $-8.2\%$ to $-20.1\%$ on Ligra CF and from $-3.8\%$ to $-12.6\%$ on GAP bc.

\observation{\label{obs:ablation-inert}Three modeled mechanisms are inert on the test set.} Removing the write-to-read floor, activation throttles, or bank-group clock changes core-cycle error by at most 0.002 percentage points; for example, the channel clock already spaces activations by one burst (8 cycles $= \max(t_{\mathrm{RRD\_S}}, t_{\mathrm{FAW}}/4)$).

\observation{\label{obs:ablation-comp}Some mechanisms compensate for each other at the core-cycle level.} Removing the last row per request type \emph{lowers} core-cycle error (1.54\% to 0.65\%) but worsens $\MAEReq$ on 22 of 24 workloads; fixing a channel-clock flaw (adding row-opening delay $\rho$ to $C$, \algref{alg:astra} line~9) lowers $\MAEReq$ on all 24 workloads (0.201 to 0.187) yet raises core-cycle error to 2.23\%. Only per-request evaluation exposes such error cancellation.

\takeaway{GPT-6 Astra's accuracy depends on per-bank state, an open-row proxy, and a write-recovery floor, while several other mechanisms are inert or cancel each other's errors.}

\subsection{Failure Mode Analysis}
\label{sec:failure-modes}
To isolate failure modes, we replay controlled synthetic traffic in an open loop (sweeping row-hit fraction, bank count, read fraction, arrival interval, FR-FCFS reordering patterns, and channel load up to saturation), reporting mean signed read-latency error normalized to oracle mean latency $L$, and check each finding on matching read classes of the 24 test workloads.

\observation{\label{obs:fm-reads}Synthesized models are accurate on read traffic below saturation.} At 30\% of oracle capacity, all four models stay within $\pm$2\% across all row-hit fractions (0--100\%) and bank counts (1--32); at 60\%, GPT-6 Astra and Opus~5.5 stay within $\pm$4\%. Under bank-group and activation-rate stress, GPT-6 Astra, Opus~5.5, and DeepSeek V4.1 Flash match the oracle exactly, and Gemini 3.8 Flash stays within $+$2.3\%.

\observation{\label{obs:fm-write-admit}Accounting for writes upon arrival over-predicts subsequent reads (Failure Mode~1).} GPT-6 Astra updates bank, bank-group, and channel clocks immediately upon write admission, whereas the oracle buffers writes. On interleaved read/write streams, GPT-6 Astra over-predicts read latency by $+$6\% to $+$88\%, while Opus~5.5 (which buffers writes) stays within $-$1\% to $+$3\%. On test workloads, the 10.3\% of reads whose bank's last admitted request was a buffered write to another row are over-predicted by 4.2 (Opus~5.5) to 43.6 cycles (DeepSeek V4.1 Flash).

\observation{\label{obs:fm-write-drain}Every model under-predicts reads delayed by write drains (Failure Mode~2).} Across the 34 synthetic points where $\ge$5\% of reads arrive during a drain, median under-prediction is 27\% for GPT-6 Astra (across 29 non-diverging points), 19\% for Gemini 3.8 Flash, and 24\% for Mess; Opus~5.5 achieves $+$0.2\% median error but diverges at 12 of 34 points. In closed-loop test runs, the 18.9\% of reads overlapping an oracle write issue are under-predicted by 5.0 (Opus~5.5) to 28.9 cycles (DeepSeek V4.1 Flash) and account for 33--47\% of total absolute per-request error, whereas reads blocked by a drain (0.22\% of test reads) account for 5--12\%. On clean reads with no oracle write delay, the other three models over-predict by 3.9--6.8 cycles, whereas Opus~5.5 over-predicts by only 0.9 cycles.

\observation{\label{obs:fm-frfcfs}No immediate-response model can predict FR-FCFS reordering, and small throughput errors amplify near saturation (Failure Modes~3 and~4).} When a row conflict arrives inside closely spaced same-bank row hits, the FR-FCFS oracle serves $k$ later hits first; every model under-predicts the conflict by $k \times t_{\mathrm{CCD\_L}}$ ($-$192 cycles for $k{=}16$) and over-predicts later hits by 86--208 cycles (though this affects only 0.08\% of single-core test reads). Near saturation, DeepSeek V4.1 Flash and Gemini 3.8 Flash sustain $\approx$6\% lower read throughput than the oracle and diverge at 95\% load, where GPT-6 Astra under-predicts read latency by a median of 14.5\%.

\takeaway{Synthesized models are accurate on read traffic below saturation; residual errors stem from writes (accounting for writes upon arrival or unpredicted drains), FR-FCFS reordering, and channel saturation.}

\section{Generalization}
\label{sec:extensions}
We evaluate how well the synthesized models and our methodology generalize beyond single-core DDR5-4800 ChampSim: without modification in gem5's O3CPU (\secref{sec:frontend-transfer}) and across nine DRAM organizations/queue depths (\secref{sec:dram-transfer}), on LPDDR6 and HBM4 both unmodified and via new synthesis runs (\secref{sec:standard-transfer}), and in preliminary multicore synthesis runs (\secref{sec:multicore}).

\subsection{A Different Core Model (gem5 O3CPU)}
\label{sec:frontend-transfer}
\begin{figure}[t]
    \centering
    \includegraphics[width=\columnwidth]{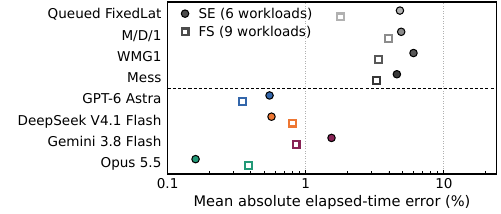}
    \caption{Mean absolute simulated elapsed-time error in gem5 O3CPU across 6 SE-mode PolyBench (circles) and 9 FS-mode NPB, GAP, and PARSEC (squares) workloads for baselines (top) and unmodified synthesized models (bottom).}
    \label{fig:extension-gem5}
\end{figure}

We evaluate all four unmodified DDR5 models integrated into gem5's execution-driven O3CPU~\cite{binkert2011gem5,lowepower2020gem5} (3.2~GHz, 32-KiB L1I/L1D, 1-MiB L2; 2M warmup, up to 2B instructions or ROI end) across 15 single-core workloads: six PolyBench~\cite{pouchet2016polybench} kernels in SE mode and nine NPB~\cite{bailey1991nas}, GAP~\cite{beamer2015gap}, and PARSEC~\cite{bienia2008parsec} benchmarks in FS mode. \figref{fig:extension-gem5} plots mean absolute simulated elapsed-time error (x-axis, log scale) separately for SE (circles) and FS (squares) workloads.

\observation{\label{obs:gem5}All four unmodified models retain their accuracy advantage in gem5's execution-driven O3CPU.}
Across all 15 workloads, Opus~5.5, GPT-6 Astra, DeepSeek V4.1 Flash, and Gemini 3.8 Flash achieve 0.30\%, 0.43\%, 0.71\%, and 1.12\% mean elapsed-time error, respectively, which are 2.7--10.1$\times$ lower than the most accurate baseline (queued FixedLat at 2.99\%, versus 3.77\% for Mess, 4.36\% for M/D/1, and 4.43\% for WMG1).

\subsection{Different DRAM Organizations and Queue Depths}
\label{sec:dram-transfer}
\begin{figure}[t]
    \centering
    \includegraphics[width=\columnwidth]{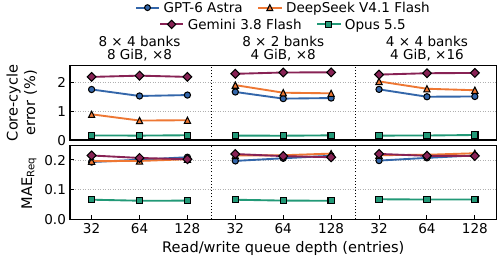}
    \caption{ChampSim test-set accuracy (core-cycle, top; $\MAEReq$, bottom) across nine DRAM bank organizations and queue depths \emph{without} modifying the models.}
    \label{fig:extension-hardware}
\end{figure}

We sweep nine hardware configurations \emph{without} modifying the models (passing only the updated hardware description): $8\times 4$, $8\times 2$, and $4\times 4$ bank groups $\times$ banks crossed with 32-, 64-, and 128-entry read/write queues on the 24 test workloads. \figref{fig:extension-hardware} plots mean core-cycle error (top) and $\MAEReq$ (bottom) across queue depths (x-axis) and bank organizations (columns).

\observation{\label{obs:dram-configs}The three models that read hardware descriptions dynamically retain accuracy across all nine configurations, while the hard-coded model degrades.}
GPT-6 Astra, Gemini 3.8 Flash, and Opus~5.5 stay within 0.25 percentage points of their synthesis-configuration error across all nine points, whereas DeepSeek V4.1 Flash's error grows from 0.68\% to up to 2.05\% on 16-bank configurations because it hard-codes the 32-bank DDR5-4800 geometry.

\subsection{Other DRAM Standards (LPDDR6 and HBM4)}
\label{sec:standard-transfer}
\begin{figure}[t]
    \centering
    \includegraphics[width=\columnwidth]{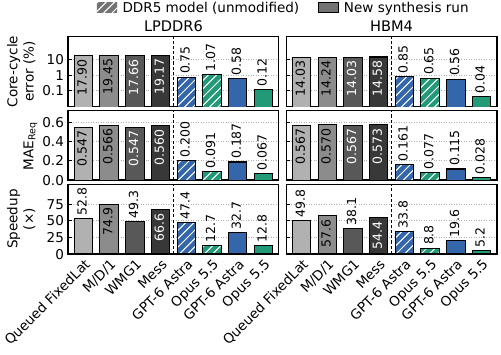}
    \caption{LPDDR6 (left) and HBM4 (right) test-set accuracy and 30-point geometric-mean speedup. Hatched: unmodified DDR5 models; solid: new ten-iteration synthesis runs.}
    \label{fig:extension-standards}
\end{figure}

We test whether our methodology generalizes to LPDDR6 (LPDDR6-10667, one $\times$12 sub-channel, 16 banks, 32-entry queues) and HBM4 (HBM4-8000, one channel with two pseudo-channels, 64 banks, 64-entry queues) using Ramulator~2's FR-FCFS oracles and the same 52-workload split (each 64-byte cache line maps to one 32-byte transfer identically for oracle and models). Baselines use each oracle's zero-load read latency and measured Mess curves. We evaluate GPT-6 Astra and Opus~5.5 both 1)~without modification via a fixed parameter-mapping adapter from DDR5, and 2)~via new ten-iteration synthesis runs from the seed model. \figref{fig:extension-standards} compares core-cycle error (top, log scale), $\MAEReq$ (middle), and 30-point geometric-mean speedup (bottom, as in \secref{sec:sc-speedup}\footnote{Measured on a newer build whose DDR5 oracle is 3.7\% faster (yielding 30.3$\times$ and 6.2$\times$ for GPT-6 Astra and Opus~5.5 on DDR5 vs.\ 31.6$\times$ and 6.4$\times$ in \figref{fig:foundation-speed}).}) for unmodified DDR5 models (hatched) and new runs (solid).

\observation{\label{obs:standards-acc}On LPDDR6 and HBM4, baselines incur 14.03--19.45\% core-cycle error, unmodified DDR5 models stay within 0.65--1.07\%, and new synthesis runs reach 0.04--0.58\%.}
We make three observations from \figref{fig:extension-standards}. First, baselines incur 17.66--19.45\% mean core-cycle error (up to 94.23\% max) on LPDDR6 and 14.03--14.58\% (up to 73.58\% max) on HBM4, with 0.547--0.573 $\MAEReq$. Second, unmodified DDR5 models achieve 0.75\% (3.22\% max) and 1.07\% (2.82\% max) on LPDDR6, and 0.85\% (4.11\% max) and 0.65\% (9.50\% max) on HBM4 (0.077--0.200 $\MAEReq$) for GPT-6 Astra and Opus~5.5, respectively; although both saturate HBM4 at half oracle bandwidth (32.0 vs.\ 62.6~GB/s) by modeling one shared bus, the test workloads' average bandwidth is at most 8.9\% of HBM4's peak. Third, new synthesis runs surpass both unmodified models and DDR5 core-cycle accuracy: on LPDDR6 and HBM4, respectively, Opus~5.5 reaches 0.12\% (0.61\% max) and 0.04\% (0.46\% max) core-cycle error (0.067 and 0.028 $\MAEReq$), while GPT-6 Astra reaches 0.58\% (2.41\% max) and 0.56\% (4.70\% max) (0.187 and 0.115 $\MAEReq$).

All synthesized models remain faster than the oracles on average: on LPDDR6 and HBM4, respectively, new GPT-6 Astra models achieve 32.7$\times$ and 19.6$\times$ speedup and Opus~5.5 achieves 12.8$\times$ and 5.2$\times$ (vs.\ 38.1--74.9$\times$ for baselines), while unmodified DDR5 models achieve 47.4$\times$ and 33.8$\times$ for GPT-6 Astra (faster than its new models) and 12.7$\times$ and 8.8$\times$ for Opus~5.5 (comparable on LPDDR6, faster on HBM4) at higher error. As on DDR5 (\secref{sec:sc-speedup}), the most accurate model is the slowest: Opus~5.5's new HBM4 model is slower than the oracle at 7 of 30 high-write/dense points (down to 0.14$\times$).

\subsection{Multicore Workloads}
\label{sec:multicore}
\begin{figure}[t]
    \centering
    \includegraphics[width=\columnwidth]{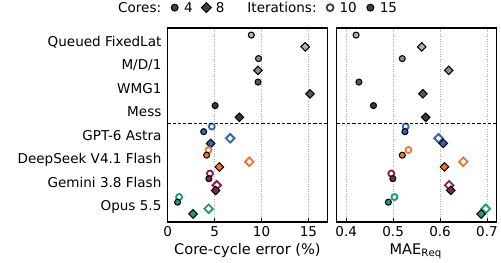}
    \caption{Multicore test-set errors over eight four-core (circles) and eight eight-core (diamonds) mixes after 10 single-core (open) and 5 multicore (filled) iterations.}
    \label{fig:extension-multicore}
\end{figure}

We run a preliminary five-iteration multicore synthesis run extending each single-core model on four- and eight-core ChampSim + Ramulator~2 mixes (eight test mixes each, 20M retired instructions per core). \figref{fig:extension-multicore} plots mean core-cycle error (left) and paired $\MAEReq$ (right) after 10 single-core (open) and 5 additional multicore (filled) iterations, and \tabref{tab:multicore-floor} compares paired $\MAEReq$ against normalized per-core Wasserstein-1 distance.

\begin{table}[t]
    \centering
    \caption{Request-level multicore accuracy on four- and eight-core mixes: paired $\MAEReq$ vs.\ normalized per-core Wasserstein-1 distance.}
    \label{tab:multicore-floor}
    \begingroup\footnotesize\setlength{\tabcolsep}{4pt}\renewcommand{\arraystretch}{1.08}
\begin{tabular*}{\columnwidth}{@{\extracolsep{\fill}}lrrrr@{}}\toprule
 & \multicolumn{2}{c}{\textbf{Paired $\MAEReq$}} & \multicolumn{2}{c}{\textbf{Wasserstein-1}} \\
\cmidrule(lr){2-3}\cmidrule(l){4-5}
\textbf{Cores} & \textbf{4} & \textbf{8} & \textbf{4} & \textbf{8} \\\midrule
Oracle, read latency +1 cycle & 0.477 & 0.683 & 0.012 & 0.009 \\
Per-core constant (median) & 0.413 & 0.547 & -- & -- \\\midrule
Queued FixedLat & 0.421 & 0.561 & 0.423 & 0.429 \\
M/D/1 & 0.519 & 0.618 & 0.506 & 0.495 \\
WMG1 & 0.427 & 0.563 & 0.479 & 0.490 \\
Mess & 0.458 & 0.569 & 0.485 & 0.498 \\\midrule
GPT-6 Astra & 0.525 & 0.607 & 0.196 & 0.245 \\
DeepSeek V4.1 Flash & 0.519 & 0.609 & 0.240 & 0.272 \\
Gemini 3.8 Flash & 0.499 & 0.623 & 0.224 & 0.298 \\
Opus~5.5 & 0.490 & 0.687 & 0.052 & 0.074 \\
\bottomrule\end{tabular*}\endgroup

\end{table}

\observation{\label{obs:mc-all}All four models achieve lower multicore core-cycle error than every baseline and 1.4--8.1$\times$ lower per-core Wasserstein-1 distance than the best baseline, whereas paired $\MAEReq$ breaks under cross-core interleaving shifts.}
On four- and eight-core mixes, respectively, all four synthesized models achieve lower core-cycle error (Opus~5.5 at 1.06\% and 2.71\%, GPT-6 Astra at 3.84\% and 4.58\%, Gemini 3.8 Flash at 4.40\% and 5.11\%, and DeepSeek V4.1 Flash at 4.16\% and 5.51\%) than the most accurate baseline on core-cycle error (Mess at 5.06\% and 7.64\%, versus 8.91--9.68\% and 9.62--15.16\% for the other three baselines). On paired $\MAEReq$, however, all four score worse than queued FixedLat, WMG1, and Mess (\tabref{tab:multicore-floor}), because any small latency difference shifts cross-core request interleaving and breaks request pairing: even the oracle shifted by $+$1 cycle scores 0.477 and 0.683, and a per-core constant median beats every model. Using the pairing-free normalized Wasserstein-1 distance (where oracle $+$1 cycle scores only 0.012 and 0.009), all four models outperform every baseline (where queued FixedLat is the best baseline at 0.423 and 0.429, versus 0.479--0.506 and 0.490--0.498 for WMG1, Mess, and M/D/1): Opus~5.5 reaches 0.052 and 0.074 (5.8--8.1$\times$ lower than queued FixedLat), and the other three models achieve 1.4--2.2$\times$ lower distance (\tabref{tab:multicore-floor}).

\takeaway{Synthesized models generalize without modification across execution-driven cores (gem5 O3CPU), DRAM organizations/queue depths (when parameters are not hard-coded), and emerging standards (LPDDR6/HBM4), while multicore synthesis motivates pairing-free request metrics.}

\section{Evaluation in a Parallel Simulator}
\label{sec:parallel-simulators}

Parallel simulators are where immediate-response models matter most (\secref{sec:bg-motivation}). We integrate the four synthesized models into zsim~\cite{sanchez2013zsim} and measure simulation speedup over a cycle-level Ramulator~2 oracle running inside zsim (focusing on speed, since end-to-end accuracy depends on simulator-specific core--memory coupling). We exclude Sniper~\cite{carlson2011sniper} because it lacks a cycle-level DRAM reference: its loosely synchronized cores issue out-of-order timestamps, and integrating a cycle-level simulator would require extensive modifications and cycle-by-cycle synchronization~\cite{eyerman2021irram}.

\head{Integration and Hardware Setup} zsim simulates 1,000-cycle intervals in a parallel \emph{bound} phase (using estimated latencies) and a serial \emph{weave} phase (simulating memory requests in timestamp order)~\cite{sanchez2013zsim}. Building on zsim's Ramulator~2 integration~\cite{esmailidokht2026integration,cpumemsimgithub}, we configure 4-GHz out-of-order cores with private 32-KiB L1I, 48-KiB L1D, and 512-KiB L2 caches, a shared 2-MiB/core L3 cache, no prefetchers, and one DDR5-4800 channel. The reference is Ramulator~2 with the oracle configuration (\secref{sec:methodology}), ticked every DRAM cycle in the weave phase; each synthesized model replaces it behind the same interface (invoked once per request at its weave event, returning an immutable departure time), sharing the same per-core exponential-moving-average bound estimator~\cite{esmailidokht2026integration}.

\begin{figure}[t]
    \centering
    \includegraphics[width=\columnwidth]{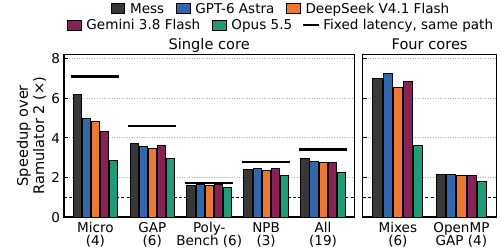}
    \caption{zsim speedup when an immediate-response model replaces cycle-ticked Ramulator~2 (geometric means; workload counts in parentheses). Horizontal lines (single-core panel) show a fixed latency through the same weave path.}
    \label{fig:zsim-speed}
\end{figure}

\head{Workloads and Measurement} We evaluate 19 single-core workloads (4 microbenchmarks, 6 GAP~\cite{beamer2015gap}, 6 PolyBench~\cite{pouchet2016polybench}, and 3 NPB~\cite{bailey1991nas} kernels; up to 200M ROI instructions, median of 2 runs) and four-core runs comprising 6 multiprogram mixes (fixed 150M-cycle window) and 4 multithreaded OpenMP GAP kernels (1 run), profiling bound plus weave time on an AMD EPYC 7B13 host pinned to 1 or 4 cores. \figref{fig:zsim-speed} shows geometric-mean speedup over Ramulator~2 across single-core (left) and four-core (right) groups; horizontal lines mark a fixed-latency model through the same weave path (zero-cost upper bound).

\observation{\label{obs:zsim-sc}With one core, synthesized models accelerate zsim by 2.2--2.8$\times$, out of at most 3.4$\times$ in this integration.} Executing identical instructions and requests, zsim runs 2.8$\times$ (GPT-6 Astra), 2.75$\times$ (DeepSeek V4.1 Flash and Gemini 3.8 Flash), and 2.2$\times$ (Opus~5.5) faster than with Ramulator~2, within 7\% of Mess (3.0$\times$) for the three faster models. Speedup rises with memory intensity, from 1.5--1.6$\times$ on PolyBench to 4.3--5.0$\times$ on microbenchmarks (whereas Opus~5.5 achieves 2.9$\times$), because Ramulator~2 ticks every DRAM cycle while immediate-response cost scales only with request count. The gap to 3.4$\times$ reflects model computation; zsim's native fixed latency (bypassing the weave phase) reaches 4.9$\times$.

\observation{\label{obs:zsim-mc}With four cores, synthesized models remove Ramulator~2's serial weave bottleneck and accelerate zsim by 3.6--7.3$\times$ (6.6--7.3$\times$ for the three faster models) on multiprogram mixes.} On four-core mixes, Ramulator~2 spends 90--95\% of runtime in the serial weave phase, capping zsim at 0.5--1.2~MIPS. Over the 150M-cycle window, synthesized models accelerate simulation by 7.3$\times$ (GPT-6 Astra), 6.9$\times$ (Gemini 3.8 Flash), 6.6$\times$ (DeepSeek V4.1 Flash), and 3.6$\times$ (Opus~5.5). Because the three faster models predict lower latencies on some requests, cores retire up to 29\% more instructions on average in the window (whereas Opus~5.5 retires 3\% fewer), yielding 7.8--8.4$\times$ (3.5$\times$ for Opus~5.5) speedup in simulated instructions per second. On multithreaded OpenMP GAP kernels (59--76\% weave time), zsim runs 2.1$\times$ faster with the three faster models and 1.8$\times$ faster with Opus~5.5.

\takeaway{Inside zsim, synthesized models remove most cycle-level DRAM overhead, accelerating simulation by 2.2--2.8$\times$ (one core) and 3.6--7.3$\times$ (four cores), with Opus~5.5 remaining the slowest.}

\section{Related Work}
\label{sec:related-work}

To our knowledge, this is the first work to use LLM agents to synthesize immediate-response DRAM latency models. \secref{sec:single-core-results} and \secref{sec:extensions} already quantitatively compare against our four closest immediate-response baselines (queued FixedLat, M/D/1, WMG1, and Mess). Below, we discuss related work in four categories.

\head{Cycle-Level and Event-Driven DRAM Simulators} Ramulator~\cite{kim2016ramulator,ramulatorgithub}, Ramulator~2~\cite{luo2024ramulator2,luo2026ramulator21,ramulator2github}, DRAMSim2~\cite{rosenfeld2011dramsim2}, and DRAMSim3~\cite{li2020dramsim3} model DRAM commands, timings, and scheduling cycle by cycle, whereas DRAMSys~\cite{steiner2022dramsys4} and gem5's DRAM controller~\cite{hansson2014simulating} use event-driven transaction-level models skipping idle cycles. Prior studies show their runtime share is modest with sequential out-of-order cores but dominant with simpler or parallel cores~\cite{li2019thesis,hansson2014simulating,sanchez2013zsim,esmailidokht2024mess}. Other works validate simulated latency against real hardware~\cite{sanchez2018missing,li2019rethinking,bostanci2026cleaning,esmailidokht2026integration}, and Esmaili-Dokht et al.~\cite{esmailidokht2026integration} propose a three-view methodology and interface enhancements resolving delayed-response mismatches when integrating Ramulator~2, DRAMSim3, and DRAMSys~4 into zsim. In this paper, we treat a cycle-level simulator as the oracle and synthesize fast immediate-response approximations matching it, complementing real-hardware calibration.

\head{Immediate-Response and Analytical DRAM Models} Fast and parallel simulators often estimate latency analytically (M/D/1 in zsim~\cite{sanchez2013zsim}, windowed M/G/1 in Sniper~\cite{carlson2011sniper}) or from bandwidth--latency curves (Mess~\cite{esmailidokht2024mess} in zsim and gem5, and PANEM~\cite{crowthers2026panem} for pre-silicon single-core flows). Because these models rely solely on aggregate load and read/write ratio, they miss row-buffer locality, bank conflicts, and write draining (\secref{sec:bg-motivation}). Recent work~\cite{baek2026address} similarly demonstrates that bandwidth--latency curves miss address-dependent channel skewness, row-buffer locality, and read/write ordering. IRRam~\cite{eyerman2021irram} manually converts Ramulator into an immediate-response model for Sniper via command histories, open-page histories, and delayed write drains. Opus~5.5 shares structural similarities with IRRam (\secref{sec:insights}), whereas GPT-6 Astra tracks per-resource free-time clocks. Concurrent work~\cite{lashkarian2026anatomy1} models each bank as an M/G/1 queue, deriving stationary mean and variance and validating its mean-latency prediction against Ramulator~2. Our work automatically synthesizes models that 1)~predict per-request latency at admission rather than stationary operating-point moments, 2)~infer row hits directly from request streams instead of taking aggregate hit/conflict probabilities as inputs, and 3)~are evaluated in a closed loop with core models on real traces.

\head{Learned and Statistical DRAM Models} Li and Jacob~\cite{li2019statistical} train decision-tree and random-forest classifiers on DRAMSim3 traces for open-loop trace replay. Feldmann et al.~\cite{feldmann2020fast} accelerate DRAMSys by augmenting its online transaction-level model with a lookup table for dense phases and training a neural-network simulator for offline trace analysis. In this paper, we synthesize readable, white-box C++ models that can be inspected and ablated (\secref{sec:model-analysis}) and evaluate them in a closed loop with core models.

\head{LLM-Driven Program Discovery} FunSearch~\cite{romeraparedes2024funsearch} and AlphaEvolve~\cite{novikov2025alphaevolve} evolve programs scored by automated evaluators. Cheng et al.~\cite{cheng2025barbarians} apply such loops to systems research, and CHIA~\cite{cui2026chia} provides the agentic hardware/software co-design framework underlying our harness (one CHIA case study aligns an existing gem5 core with RTL simulation, whereas our agents synthesize a new memory model from scratch). In computer architecture, LLM agents and evolutionary search design cache replacement policies~\cite{gupta2026archagent,mhapsekar2026cacheforge,kypriotis2026agentic}, prefetchers~\cite{gonzalez2026archagentv2,kypriotis2026agentic}, and branch predictors~\cite{kypriotis2026agentic}, and explore microarchitecture design spaces~\cite{xiong2026microevo}, scoring designs by performance (e.g., IPC) or PPA. Rosetta~\cite{sankaralingam2026rosetta} uses a multi-agent LLM pipeline to derive analytical performance models from paper PDFs without a simulator oracle. ArchEval~\cite{wang2026archeval} and PerfReasoning~\cite{zhao2026perfreasoning} benchmark LLM agents and LLMs on architecture design, performance reasoning, and analytical model generation, while LLMulator~\cite{chang2025llmulator} fine-tunes an LLM directly as a black-box cost model for dataflow accelerators. Our work targets white-box memory simulation by evaluating candidates on complementary core-cycle and per-request metrics, withholding a hidden test set, and analyzing synthesized mechanisms and failure modes.

\section{Conclusion}
\label{sec:conclusion}

We propose an agentic methodology in which LLM agents autonomously synthesize fast, accurate, white-box immediate-response DRAM latency models. Using a closed-loop development harness around Ramulator~2 and ChampSim that evaluates both core-cycle error and normalized per-request error ($\MAEReq$), we conduct four ten-iteration synthesis runs across four LLMs. Every synthesized model outperforms all four baselines on both metrics: Opus~5.5 achieves 0.15\% mean core-cycle error and 0.063 $\MAEReq$ (\CoreReduction$\times$ and \RequestReduction$\times$ lower than Mess and queued FixedLat, respectively) at 6.4$\times$ DRAM speedup, while GPT-6 Astra achieves 1.54\% error and 0.201 $\MAEReq$ at 31.6$\times$ speedup. Without modification, the models retain accuracy in gem5 O3CPU (0.30--1.12\% vs.\ 2.99--4.43\%) and, for three models, across nine DRAM configurations. On LPDDR6 and HBM4, new synthesis runs reach 0.04--0.58\% error (vs.\ $\ge$14.03\% for baselines). In zsim, synthesized models accelerate four-core multiprogram simulation by 3.6--7.3$\times$ (2.2--2.8$\times$ single-core). White-box analysis reveals a few interpretable mechanisms and identifies writes as a major remaining error source.

We hope our methodology, findings, and synthesized models inspire future work on fast, accurate, and auditable simulation models for other system components. We will open-source our harness, models, and evaluation infrastructure.

\section*{Generative AI Usage Acknowledgment}
Besides their role as agents studied in this paper, we used GPT-6 Astra and Opus~5.5 for 1)~providing generic coding assistance, 2)~streamlining and automating experiment workflows, 3)~drafting figures and plots, and 4)~editing and drafting manuscript text. The authors reviewed all generated content and take full responsibility for the paper.

\bibliographystyle{unsrt}
\bibliography{references}

\end{document}